# A Global Radio Remote Sensing Network for Observing Space Weather Dynamics

Ryan Volz, Philip J. Erickson | MIT Haystack Observatory
Jorge L. Chau | Leibniz Institute of Atmospheric Physics at the University of Rostock
Thomas Y. Chen | Columbia University
Scott E. Palo | University of Colorado Boulder
Juha Vierinen | UiT Arctic University of Norway

**Introduction:** The need for more and better data is a constant refrain in geospace science. Past workshops and decadal surveys have noted a compelling need for real-time observations and a clear realization that our current sampling of the vast geospace environment is wholly insufficient to measure the highly variable (both in space and time) space environment.[1] Regular, densely-sampled measurements, even if they lack the detail of those from our flagship instruments, would be a boon to scientific understanding and modeling of the geospace system. These ideas are not new, but we posit that the technology has now arrived to make dense observations of the upper atmosphere feasible in cost and effort. This white paper sketches out the scientific rationale for a network of radio instruments delivering dense observations of the near-Earth space environment and the broad steps necessary to implement wide-scale coverage in the next 30 years.

**Scientific rationale:** Space weather provides the background quiet time climatology and geomagnetic storm time conditions that must be predicted and accommodated for success of both future technological systems and human presence in the near-Earth space environment. However, our observational knowledge of the space environment is decades behind that of lower atmosphere terrestrial weather for many reasons. Chief among these is the difficulty in enabling sufficient remote sensing observations, due to the physics-imposed requirement to sample vast spatial and temporal scales. Factors such as charged particle dynamics experiencing electrodynamic forcing over very long distances, and corresponding two-way coupled influences on the neutral upper atmosphere, pose a considerable challenge to data collection and understanding.

Our ability to predict the terrestrial weather has continuously improved and provides a useful comparative case study of what is possible with sufficiently dense observations and appropriately driven models. Consider that currently a 5-day forecast is as accurate as a 24-hour forecast was in 1980, and long-term forecasts of a week or more are now useful. On the path to better forecast success, improvements in observations have been a critical element, both in supplying real-time data for assimilation but also in amassing archival data that can be used for post-analysis, model skill assessment, and improvement of the key physical processes included in the models. The data from radiosondes launching every 12 hours for 365 days a year and measuring winds, temperature, pressure, and relative humidity, along with continuous radar data on winds and precipitation, make this forecast capability a reality. In contrast, the upper atmosphere and in particular the mesosphere-lower thermosphere (MLT) system is far less densely sampled. This latter fact is in contrast to the vast number of complex and interesting frontier science topics that have yet to be clarified, such as atmospheric gravity wave breaking and momentum transfer, severe neutral wind shear effects on ionospheric E region variability, and ionosphere-thermosphere mass and energy dynamics. Despite this fact, unlocking the secrets of the intimately coupled physical pathways in the upper neutral and ionized atmosphere has critical importance for studying whole atmosphere physics, and these subjects are at the frontier of understanding the space environment.

Particularly important ionosphere-thermosphere science topics in need of significant additional study include the influence of lower atmospheric disturbances on the upper atmosphere. Specifically, estimated wave energy flux in the lower thermosphere ("space weather from below"[2]) is comparable to daily average Joule power input from above to the thermosphere[3]. Furthermore, this wave flux is likely underestimated by a large amount since most model simulations do not yet capture the full wave spectrum. Additionally, little is still known about the relative importance of planetary waves, tides, and gravity waves taken as a joint whole, and how these influences evolve in both time and space. Appropriate coupled analysis in fact is largely absent from the literature today due to a lack of sufficiently dense observations, and so typical studies have focused only on isolated aspects such as tidal forcing.[4] A similar set of observational statements can be made for specification of the full atmospheric wave spectrum (time periods, spatial wavelengths) in the ionosphere-thermosphere system, with a goal of determining how these regions are structured and how and when the system selects preferential wavelengths exhibiting high coupling efficiency.

Additionally, the current state of the art whole atmosphere coupling models remain relatively poor at propagating input energy influences throughout the ionosphere-thermosphere system when run in an observationally unconstrained mode. For example, Pedatella et al. (2014)[5] shows that four different modeling runs without data constraints above ~50 km altitude have fundamentally inconsistent responses in the mesosphere-lower thermosphere system. These large discrepancies in predicted response then exhibit a further and similarly large uncertainty in projected ionospheric influences and ionosphere-thermosphere energy exchange.[6] As one would expect, data assimilation-driven models show clear improvement in predicted vs. observed response when mesosphere-lower thermosphere observations are used as compared to a free-running prediction absent these data.[7]

Observationally, MLT data from the TIMED satellite has been shown in the literature to be effective (when available) at improving model performance and prediction. However, TIMED has been on orbit since 2001 and its ~95-minute orbital period presents a challenge to wave interpretation. The satellite is only in one location at any given time, which makes it impossible to measure the spatial/temporal variability of the MLT as this variability is not global nor shorter than a period of 1 day. Clearly, the important mesoscale structure which drives space weather systems is severely undersampled. By contrast, consider the lower atmosphere observing strategy which consists of both low earth orbit and geostationary satellite observations coupled with extensive ground based observations of the system.

There is a clear need for more modern observations of the near-Earth space environment with better local time coverage. In particular, a ground-based sensor locked to the planet's rotation provides a badly needed and comprehensive view of MLT conditions and wave activity near orographic gravity wave generating features such as mountains. Ground-based sensors also have the considerable advantage of resolving short-term temporal variations in the MLT wave spectrum compared to e.g. TIMED data, which can only provide averaged information over long spatial and temporal scales that obscure vitally important dynamics. If we take a page from the terrestrial weather community, it is clear that a significant investment in a real-time mesoscale observing network is critical to both advance community understanding of the near-Earth space region and its role in the whole atmosphere system and to unlock the development of a robust space weather forecast system.

**Supporting measurement techniques:** A dense and scalable network of ground-based remote sensing instruments can be composed of a variety of measurement techniques. Each observes a slice of the fundamental parameters of interest, including neutral winds and ionospheric density, to collectively describe the physical processes in the near-Earth space environment. We highlight two of the key enabling technologies here, although we note that other techniques and instruments can play a large role in providing complementary measurements (e.g. HF amateur radio beacons used for tomography of ionospheric irregularities).

*MIMO meteor radar [MLT wind field]*: Recent developments in multiple-input multiple-output (MIMO) meteor radar networks have made higher-resolution wind measurements of the upper atmosphere possible[8]. These networks operate over coded continuous-wave links between separately-located transmitter and receiver sites to increase the density of specular meteor trail observations and provide diversity in sensing Doppler-derived wind projections. Such datasets contain enough information to estimate the three-dimensional wind field within the observation volume to a resolution limited only by the measurement density in space and time. *Low-power ionosonde [bottomside ionospheric density]*: Coded continuous-wave transmissions also enable a cross-linked network of low-power ionosondes operating in much the same manner as the above-described MIMO meteor radar network. Each transmitter and receiver pair can be used to produce an oblique ionogram. With combinatorial scaling, this provides a cost-effective way to densely sample the ionospheric density along each TX-RX link. Innovations such as the electro-magnetic vector sensor (6 orthogonal antenna elements with a common phase center) could help to increase each individual instrument's degrees of freedom further. This could lead to volumetric imaging of the bottom side ionosphere within the regional network.

**Network composition:** The envisioned network would consist of nodes consisting of at least the above instruments. A key to feasibility is that required infrastructure, consisting of site power, high-bandwidth internet, licensing, and/or building and ground space, can be co-located at relatively few sites. These anchor sites would thus encompass meteor radar and ionosonde transmitters, meteor radar interferometric antenna arrays, and computational resources. Receiver sites would require comparatively little infrastructure and can be operated off-grid with solar power and/or wireless internet, so they have much greater freedom to be located where necessary and in greater numbers to fill out the network. The cross-linked radar nodes form the backbone of the network, and other instruments (including but not limited to GNSS receivers) are envisioned to piggy-back onto that infrastructure investment with relatively low add-on cost. By focusing on low cost and modest power, space, and connectivity requirements for the majority of the network sites, it becomes possible to engage with educators and amateur enthusiasts to expand and support the network. Community-focused receiver sites can be built with commodity equipment and located on rooftops or in backyards, and feeding data back into the network can be enabled with open source software. HamSCI[9] has similarly engaged the amateur radio community to much success, and this would be a way to expand and strengthen those connections.

**Progression timeline:** Since the network of radio instruments described above can be expanded simply by adding more nodes, we suggest a staged deployment beginning with targeted dense regional deployments and expanding with maturing technology to eventually span continents or even the globe.

*Current state (2020)*: Efforts are already underway to deploy regional MIMO meteor radar networks and develop the low-power ionosonde into an operational instrument. GNSS TEC receivers already cover much of the globe, but density can always be improved. These instruments/deployments are being developed and funded separately, and near-term progress will be through developing the supporting technologies and investing in dense regional networks that complement existing flagship instruments.

*2025*: By this time, we expect the currently-planned regional networks to be operational and planning for the next stage of expansion to be well underway. Expansion will necessarily involve overlapping the coverage areas of the supporting instruments and combining future deployments. These regional networks will individually allow for novel study of mesoscale features with regional connections (e.g. the influence of orography or overhead convection) and collectively add coverage for global studies (e.g. tides or planetary waves). Even at this regional scale, the scientific return would be significant.

*2030-2035*: The primary objective for this timeframe is to achieve large-region or continental-scale dense coverage with a network composed of all of the complementary instruments. Such an effort will require leadership from a core group with significant community backing and investment. Long-term planning will have to start now to reach that goal. At this stage scientifically, the network will unlock temporal and spatial variations in mesoscale features and provide enough regular observations for data assimilation models to drive significant improvement over the current state of the art.

*2050*: The ultimate goal is to span the globe with a network of radio instruments that can observe the physical processes in the near-Earth space environment with enough density in time and space so as to revolutionize our scientific understanding. The supporting technologies will be mature enough at this point that expanding coverage to new regions will be a question only of scientific utility.